\documentclass[12pt,preprint]{aastex}
\shorttitle{ Evolution of the Reverse Shock Emission from SNR 1987A}
\shortauthors{Heng et al.}
\begin{document}

\title{EVOLUTION OF THE REVERSE SHOCK EMISSION FROM SNR 1987A}

\author{Kevin Heng\altaffilmark{1}, Richard McCray\altaffilmark{1}, 
Svetozar A.\ Zhekov\altaffilmark{2}, Peter M.\ Challis\altaffilmark{3},
Roger A. Chevalier\altaffilmark{4}, Arlin P.S. Crotts\altaffilmark{5}, Claes Fransson\altaffilmark{6}, Peter Garnavich\altaffilmark{7}, Robert P. Kirshner \altaffilmark{3}, Stephen S. Lawrence\altaffilmark{8}, Peter Lundqvist\altaffilmark{6}, Nino Panagia\altaffilmark{9}, C.S.J. Pun\altaffilmark{10}, Nathan Smith \altaffilmark{11}, Jesper Sollerman\altaffilmark{12}, Lifan Wang\altaffilmark{13}}

\altaffiltext{1}{JILA, University of Colorado, 440 UCB, Boulder, CO
80309-0440; hengk@colorado.edu, dick@jila.colorado.edu}

\altaffiltext{2}{Space Research Institute, Moskovska str. 6, Sofia-1000, Bulgaria}

\altaffiltext{3}{Harvard-Smithsonian Center for Astrophysics, 60 Garden Street, Cambridge, MA 02138}

\altaffiltext{4}{Department of Astronomy, University of Virginia, P.O. Box 3818, Charlottesville, VA 22903-0818}

\altaffiltext{5}{Columbia Astrophysics Laboratory, Columbia University, 550 West 120th Street, New York, NY 10027}

\altaffiltext{6}{Stockholm Observatory, Department of Astronomy, AlbaNova, SE-106 91, Stockholm, Sweden}

\altaffiltext{7}{Department of Physics, University of Notre Dame, 225 Nieuwland Science Hall, Notre Dame, IN 46556}

\altaffiltext{8}{Department of Physics and Astronomy, 151 Hofstra University, Hempstead, NY 11590}

\altaffiltext{9}{Space Telescope Science Institute, 3700 San Martin Drive, Baltimore, MD 21218}

\altaffiltext{10}{Department of Physics, University of Hong Kong, Pokfulam Road, Hong Kong, China}

\altaffiltext{11}{Center for Astrophysics and Space Astronomy, University of Colorado, 389 UCB, Boulder, CO 80309; Hubble Fellow}

\altaffiltext{12}{Dark Cosmology Center, Niels Bohr Institute, University of Copenhagen, Juliane Maries Vej 30, DK-2100 Copenhagen \O, Denmark}

\altaffiltext{13}{Institute for Nuclear and Particle Astrophysics, E.O. Lawrence Berkeley National Laboratory, Berkeley, CA 94720}

\begin{abstract}
We present new (2004 July) G750L and G140L Space Telescope Imaging Spectrograph (STIS) data of the H$\alpha$ and Ly$\alpha$ emission from supernova remnant (SNR) 1987A.  With the aid of earlier data, from Oct 1997 to Oct 2002, we track the local evolution of Ly$\alpha$ emission and both the local and global evolution of H$\alpha$ emission.  We find that the average H$\alpha$ intensity has increased locally by a factor $\sim$ 3 for both blueshifted and redshifted emission, over periods of about 5 and 7 years, respectively.  The average Ly$\alpha$ intensity has increased by a factor $\sim$ 9, over about 5 years, for both components.  The most recent observations allow us, for the first time, to directly compare the H$\alpha$ and Ly$\alpha$ emission from the same slit position and at the same epoch.  Consequently, we find clear evidence that, unlike H$\alpha$, Ly$\alpha$ is reflected from the debris by resonant scattering.  In addition to emission which we can clearly attribute to the surface of the reverse shock, we also measure comparable emission, in both H$\alpha$ and Ly$\alpha$, which appears to emerge from supernova debris interior to the surface.  New observations taken through slits positioned slightly eastward and westward of a central slit show a departure from cylindrical symmetry in the H$\alpha$ surface emission.  No obvious asymmetry is seen in the interior emission.  Using a combination of old and new observations, we construct a light curve of the total H$\alpha$ flux, $F$, from the reverse shock, which has increased by a factor $\sim$ 4 over about 8 years.  However, due to large systematic uncertainties, we are unable to discern between the two limiting behaviours of the flux $-$ $F \propto t$ (self-similar expansion) and $F \propto t^5$ (halting of the reverse shock).  Such a determination is important for constraining the rate of hydrogen atoms crossing the shock, which is relevant to the question of whether the reverse shock emission will vanish in $\lesssim 7$ years (Smith et al. 2005).  Future deep, low- or moderate-resolution spectra are essential for accomplishing this task.
\end{abstract}

\keywords{circumstellar matter --- shock waves --- supernovae:
individual (SN1987A) --- supernova remnants}

\section{INTRODUCTION}

SNR 1987A serves as a unique astrophysical laboratory for the study of the physics of shocks.  Prior to the supernova (SN) explosion, the progenitor, Sanduleak $-$69\arcdeg 202 (Type B3 I blue supergiant; Rousseau et al. 1978), was surrounded by a circumstellar ring system, consisting of an inner equatorial ring and two outer rings (Cassatella et al. 1987; Fransson et al. 1987; Crotts, Kunkel \& McCarthy 1989; Crotts \& Kunkel 1991; Wang \& Wampler 1992; Crotts, Kunkel \& Heathcote 1995).  The impact of the ejecta with the circumstellar rings has been anticipated since the discovery of the circumstellar gas and various predictions had been made regarding the time of impact (Luo \& McCray 1991; Luo, McCray \& Slavin 1994; Chevalier \& Dwarkadas 1995).  As the blast wave propagates through the ambient medium, a double shock structure is established (Chevalier 1982).  A shocked H~{\sc ii} region resides behind the blast wave, and a contact discontinuity separates this region from the shocked ejecta.  H$\alpha$ and Ly$\alpha$ emission from the reverse shock result from electron and ion impact excitation of neutral hydrogen atoms that cross the reverse shock and enter the shocked ejecta (Borkowski, Blondin \& McCray 1997, hereafter B97).  As the hydrogen atoms are freely streaming, there is a unique mapping of the Doppler shift, $\Delta\nu$, of the emitted photons to the depth of the supernova debris: 
\begin{equation}
\frac{\Delta\nu}{\nu_0} = \frac{z}{ct},
\label{eq:doppler}
\end{equation}
\noindent
where $\nu_0$ is the initial frequency of the photon, the depth, $z$, is measured from the midplane of the debris along the line of sight, and $t$ is the time since the supernova explosion.  Since the equatorial ring is at an inclination of $\sim 42 \arcdeg - 44\arcdeg$ (Panagia et al. 1991; Burrows et al. 1995; Plait et al. 1995; Sugerman et al. 2002), the blueshifted and redshifted emission are associated with the northern and southern sides of the debris, respectively.  The emission mechanism is similar to the one present in Tycho's remnant (Chevalier \& Raymond 1978).

Since the work of Sonneborn et al. (1998, hereafter S98), the SAINTS (Supernova 1987A INTensive Study) group has been observing the reverse shock emission with the Space Telescope Imaging Spectrograph (STIS) onboard the {\it Hubble Space Telescope (HST)}.  S98 detected the first definitive sign of impact between the blast wave and the equatorial ring, in 1997 April STIS spectral images in Ly$\alpha$.  Michael et al. (1998a) built a simple kinematic model, based the data obtained by S98, and found that the high-velocity Ly$\alpha$ emission could be explained by $\sim$ 15,000 km s$^{-1}$ hydrogen atoms crossing a reverse shock in the shape of a slightly prolate ellipsoid.  Subsequently, Michael et al. (1998b) found that a model in which the reverse shock emission was confined primarily to an equatorial band, as proposed by B97, agreed better with new (1997 Sept \& Oct) data.  They also detected the presence of residual Ly$\alpha$ emission that appeared to come from a volume interior to the reverse shock surface, and a departure from axisymmetry in both H$\alpha$ and Ly$\alpha$ emission, which correlated with observations of non-thermal radio emission (Gaensler et al. 1997).  Michael (2000) and Michael et al. (2000) followed up with more STIS observations and proposed that ``hot spots''  are formed where the blast wave overtakes density protrusions on the equatorial ring.  Pun et al. (2002) analyzed and modeled the optical and ultraviolet spectrum of the first hot spot, also known as ``Spot 1'' and found that the emission from the shocked gas came from slower ($\lesssim 135$ km s$^{-1}$), radiative shocks.  Lawrence et al. (2000) and Sugerman et al. (2002) analyzed the emergence of several more hot spots.  Michael et al. (2003, hereafter M03), analyzed high-velocity ($\sim \pm 12,000$ km s$^{-1}$) H$\alpha$ and Ly$\alpha$ emission profiles to map the geometry and development of the reverse-shock surface and found its shape to be consistent with a model in which the supernova debris expanded into a bipolar nebula.  M03 also found evidence for the resonant scattering of Ly$\alpha$ photons within the supernova debris and detected emission in both H$\alpha$ and Ly$\alpha$ that appeared to come from inside the reverse shock surface.

With the demise of STIS, it is an appropriate time to consolidate what we have learnt about the reverse shock emission and its evolution.  In \S\ref{sect:reduction}, we list the observations and describe our data reduction procedures.  In \S\ref{sect:local}, we analyze the evolution of the H$\alpha$ and Ly$\alpha$ emission from the reverse shock.  Because telescope time was limited, we were unable to completely map the emission at each epoch.  Therefore, the observations are sparse and we can directly compare observations at different epochs only in restricted regions where the observations overlap.  In our most recent (July 2004) set of observations with STIS, we obtained spectra of the reverse shock in both H$\alpha$ and Ly$\alpha$ with the same slit.  We compare these spectra in \S\ref{sect:compare} and \S\ref{sect:interior}.  In \S\ref{sect:Halphalightcurve}, we construct the light curve of the reverse shock H$\alpha$ emission from new (July 2004) and old (Oct 1997 to Oct 2002) STIS observations.  We also include ground-based data from Smith et al. (2005, hereafter S05), obtained during the commissioning run of the Low Dispersion Survey Spectrograph-3 (LDSS3) on the Clay Telescope of the Magellan Observatory.  In \S\ref{sect:discussion} and \S\ref{sect:summary}, we interpret and summarize our results. 

\section{OBSERVATIONS \& DATA REDUCTION}
\label{sect:reduction}

\begin{figure}
\begin{center}
\includegraphics[width=6.5in]{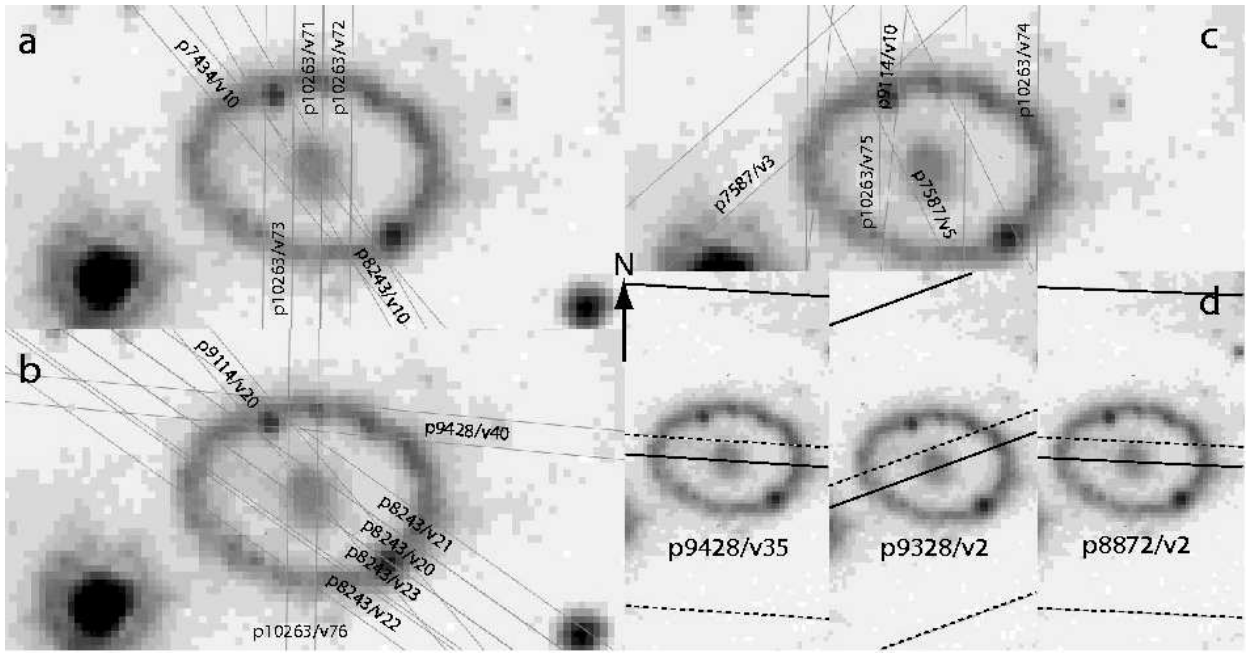}
\end{center}
\caption{Slit overlays for the STIS observations.  (a) 0$\farcs$2 G750L. (b) 0$\farcs$2 G140L.  (c) 0$\farcs$5 G750L.  Only 
p10263/v74 and p10263/v75 are used in our analysis.  (d) 2$^{\prime\prime}$ G750L.}
\label{fig:overlays}
\end{figure}

We analyze STIS observations of SNR 1987A, taken with the G750L and G140L gratings, from 1997 Oct 6 to 2004 July 23 (days 3878 to 6360 since the SN explosion).  In Table~\ref{tab:observations}, we list these observations (``pX/vY'' means ``Proposal X, Visit Y'') and in Fig.~\ref{fig:overlays}, we display the corresponding slit positions overlaid on images of the circumstellar ring.  All velocity scales are relative to the systematic velocity of SNR 1987A, $v_{87A}=289$ km s$^{-1}$ (Crotts \& Heathcote 2000).

\subsection{CLEANING}
\label{sect:cleaning}

We employ standard techniques to clean the data, as described by M03.  Each observation consists of three or four exposures taken at dithered positions along the slit.  The spectral resolutions are $\sim$ 450 km s$^{-1}$ and $\sim$ 300 km s$^{-1}$ for the G750L and G140L data, respectively.  We reduce the two-dimensional spectra using standard calibration files.  We remove hot pixels by utilizing the {\it cr\_reject} procedure first written by Hill et al. (1997), using a user-supplied error cube and default settings for the clipping sigmas.  We then combine the exposures.  The resulting images (Fig.~\ref{fig:rawimages}) contain not only spectra of the reverse shock but also a number of contaminants as indicated.

In the G750L spectra, we see emission lines from the circumstellar ring at H$\alpha$, [N~{\sc ii}] $\lambda\lambda$6548, 6584, [O~{\sc i}] $\lambda\lambda$6300, 6364, [S~{\sc ii}] $\lambda\lambda$6717, 6731, and He~{\sc i} $\lambda$6678.  These are due to residual recombination and cooling following the initial ionizing flash of the supernova (Fransson \& Lundqvist 1989).  We also see medium-velocity ($\sim \pm 3000$ km s$^{-1}$) emission from the radioactive core (defined as the central ejecta powered by the decay of radioactive species) of the supernova in H$\alpha$ and [O~{\sc i}] $\lambda\lambda$6300, 6364. We mask out these contaminants and remove both background emission and stellar continuua using average profiles.  We deredden the spectra using the extinction curve towards Star 2 (Scuderi et al. 1996).  

In the G140L spectra the main contaminants are N~{\sc v} $\lambda\lambda$1239, 1243 emission, broad damping wings from interstellar H~{\sc i} absorption, geocoronal emission at Ly$\alpha$ and a few fainter lines, and interstellar line absorption due to C~{\sc i} $\lambda\lambda$1191, 1259 and N~{\sc i} $\lambda$1199.  We remove the geocoronal emission by subtracting an average spectrum and then masking out the central slit.  We mask out the N~{\sc v} $\lambda\lambda$1239, 1243 emission, since its intrinsic distribution from the ring system is unknown.  We construct composite absorption profiles by spatially integrating and combining spectra from many separate observations and use these to correct for the interstellar absorption lines.  To correct for broad damping wings due to interstellar Ly$\alpha$ absorption, we divide the data by an absorption profile corresponding to $N_{HI} = 3 \times 10^{21}$ cm$^{-2}$, following M03.  In all of the observations, the statistical errors are negligible ($\sim 0.1 \%$).

\subsection{COMPONENTS OF THE REVERSE SHOCK EMISSION}

\begin{figure}
\begin{center}
\includegraphics[width=5in]{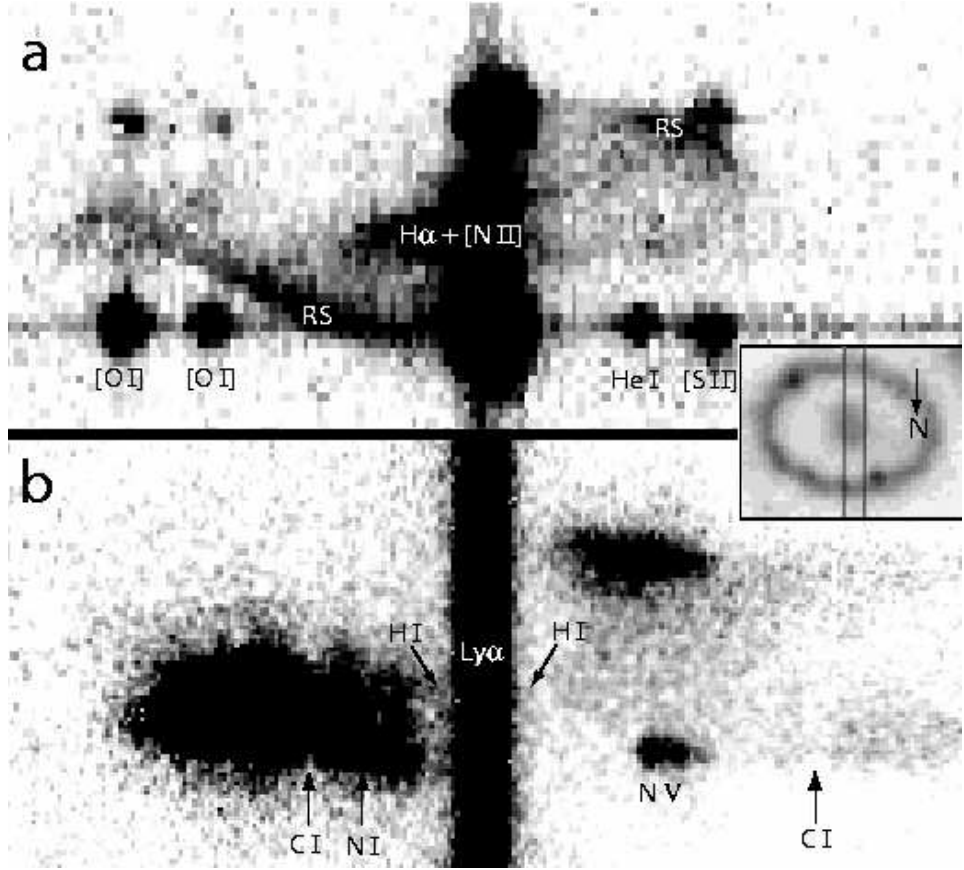}
\end{center}
\caption{(a) Sample G750L observation (p10263/v71) with the following contaminants indicated:  The vertical stripe is due to emission 
from interstellar H$\alpha$ and [N~{\sc ii}] $\lambda\lambda$6548, 6584.  Vertically-aligned pairs of spots are emission from the 
circumstellar ring at H$\alpha$, [N~{\sc ii}] $\lambda\lambda$6548, 6584, [O~{\sc i}] $\lambda\lambda$6300, 6364, [S~{\sc ii}] 
$\lambda\lambda$6717, 6731, and He~{\sc i} $\lambda$ 6678.  The horizontal feature located midway between the spots is H$\alpha$ 
emission from the radioactive core of the supernova.  A fainter such feature due to core emission in [O~{\sc i}] $\lambda\lambda$6300, 
6364 is also evident.  (b) Sample G140L observation (p10263/v76) with the following contaminants indicated: N~{\sc v} $\lambda\lambda$1239, 1243 emission from a hot spot; broad damping wings from interstellar H~{\sc i} absorption; geocoronal Ly$\alpha$ emission; interstellar line absorption due to C~{\sc i} $\lambda\lambda$1191, 1259 and 
N~{\sc i} $\lambda$1199.  In both cases, the insert shows the slit position overlaid on an image of the circumstellar ring.}
\label{fig:rawimages}
\end{figure}

The H$\alpha$ emission streaks labeled {\bf RS} in the G750L observation in Fig.~\ref{fig:rawimages}a are evidence that the slit is intersecting a curved surface of the reverse shock, on which the emitting hydrogen atoms have a unique velocity at every position.  We call these features ``surface emission''.  In addition to these streaks, one can also see faint H$\alpha$ emission extending to lower velocity at the same vertical positions.  If this emission comes from freely-streaming hydrogen atoms in the supernova debris, then it must originate from within the reverse shock surface.  Therefore, we call it ``interior emission''.  When we refer to the ``reverse shock emission'', we mean both the surface and the interior emission (see \S\ref{sect:interior} and \S\ref{sect:interior2}). Wherever possible, we isolate the former from the latter.  We emphasize that the interior emission is distinct from the core emission (see \S\ref{sect:cleaning}).  In the G140L Ly$\alpha$ spectra (e.g., Fig.~\ref{fig:rawimages}b), the interior emission is so prominent that it is difficult to distinguish from the surface emission.  As noted by M03, the interior emission in H$\alpha$ is not subject to resonant scattering, as is the case for Ly$\alpha$.  Therefore, we use the H$\alpha$ surface emission streaks as templates to locate the surface emission in the Ly$\alpha$ spectrum (see \S\ref{sect:compare}). 

\subsection{SPATIAL-SPECTRAL CORRECTION}

\begin{figure}
\begin{center}
\includegraphics[width=4.5in]{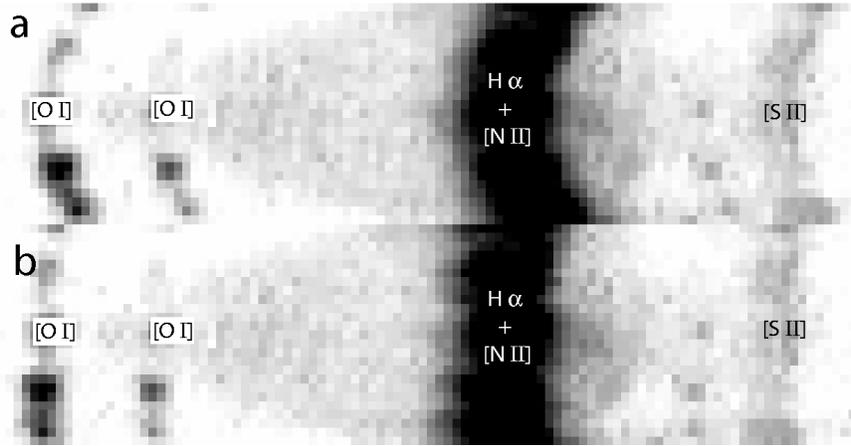}
\end{center}
\caption{Sample 2$^{\prime\prime}$ observation (northern half of p9428/v35).  (a) Original spectrum.  (b) Spectrum after 
spatial-spectral correction is applied.  The procedure uses the [O~{\sc i}] $\lambda$6300 line as a reference.}
\label{fig:sample}
\end{figure}

While the emission from the SN core fades, that from the reverse shock brightens.  Since the G750L 2$^{\prime\prime}$ observations were taken at earlier epochs, reducing them is more challenging than for the later observations, due to the low signal-to-noise ratio and the presence of multiple contaminants.  Furthermore, in STIS observations taken with the 2$^{\prime\prime}$ slit, each point along a vertical (i.e., spatial) cut corresponds to a different velocity, owing to the curved shape of the reverse shock surface viewed through the slit.   To correct for this curvature, we apply a spatial-spectral correction to morph each feature into a ``bar'', using the [O~{\sc i}] $\lambda$6300 line from the unshocked ring as a reference (Fig.~\ref{fig:sample}).  This procedure not only facilitates the masking out of contaminants, but also ensures that the spatially summed data are placed in the correct velocity bins.  The correction is applied for the 0$\farcs$5 observations as well.

\section{LOCAL EVOLUTION OF REVERSE SHOCK EMISSION}
\label{sect:local}

\begin{figure}
\begin{center}
\includegraphics[width=5.5in]{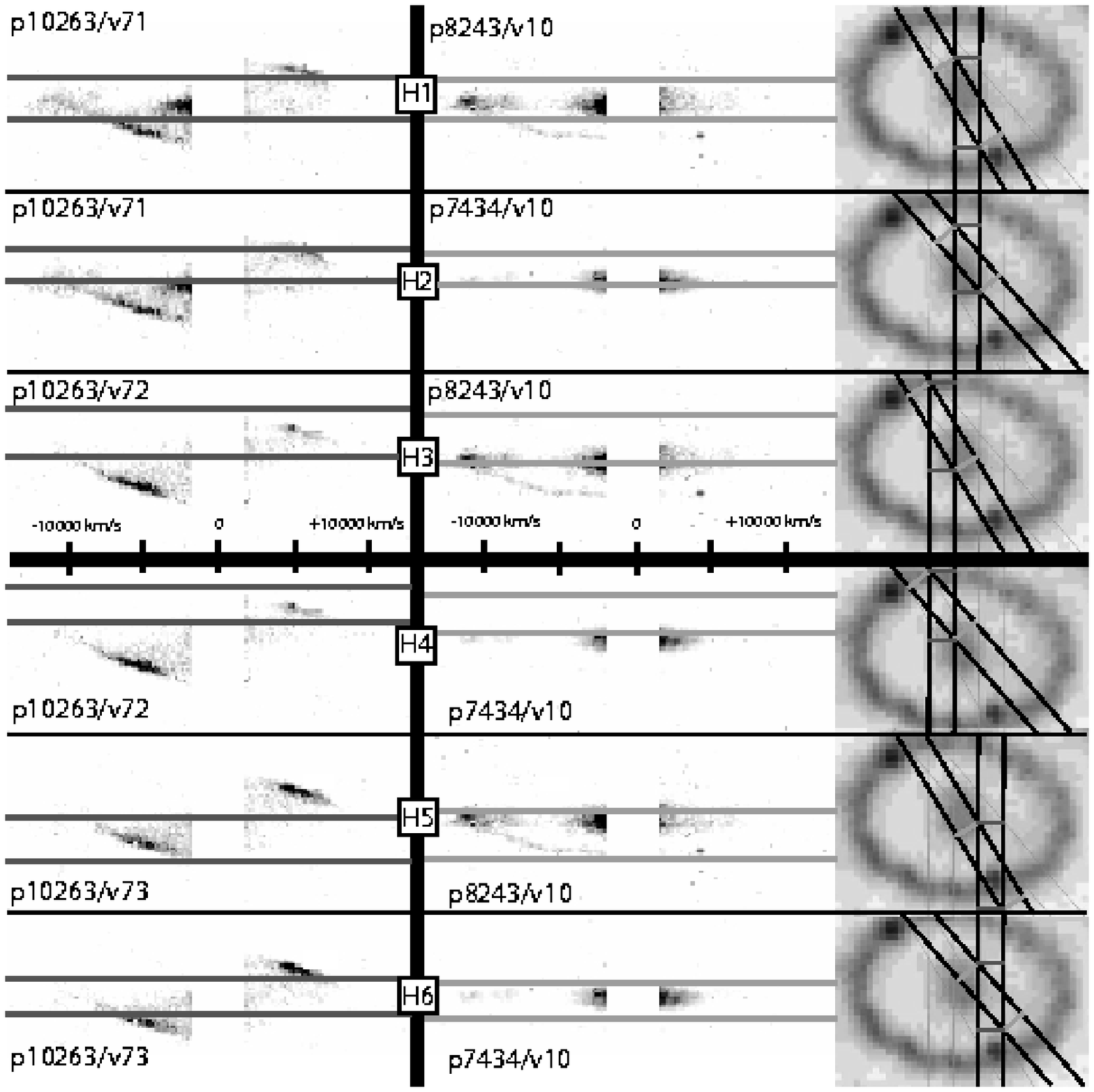}
\end{center}
\caption{Montage of the 0$\farcs$2 G750L observations used to track the local evolution of H$\alpha$ emission.  Each pair of 
observations is labeled as shown, with the newer (2004) observation being displayed in the left column; the corresponding slit 
overlays are in the right column.  In all of the slit overlay panels, North is down.  The newer (2004) observation always has its slit 
position aligned parallel to the North-South line.  The region of overlap is bracketed by the lines shown.  There are four points of overlap and summing between the innermost or outermost ones, in the vertical direction, has a negligible effect on the intensity ratios obtained (right panel of Fig.~\ref{fig:local_h_alpha}).}
\label{fig:montage_h_alpha}
\end{figure}

\begin{figure}
\begin{center}
\includegraphics[width=5in]{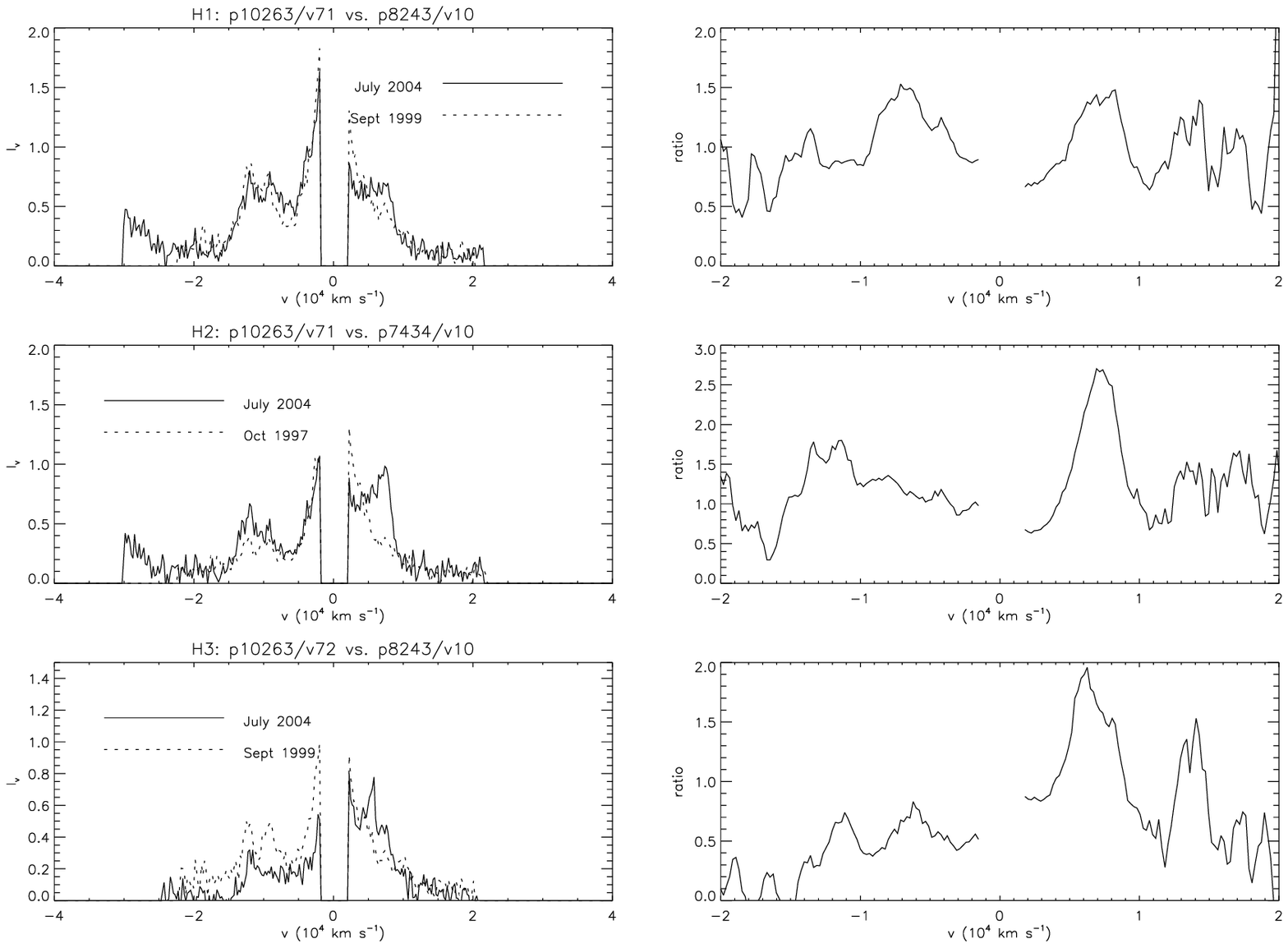}
\includegraphics[width=5in]{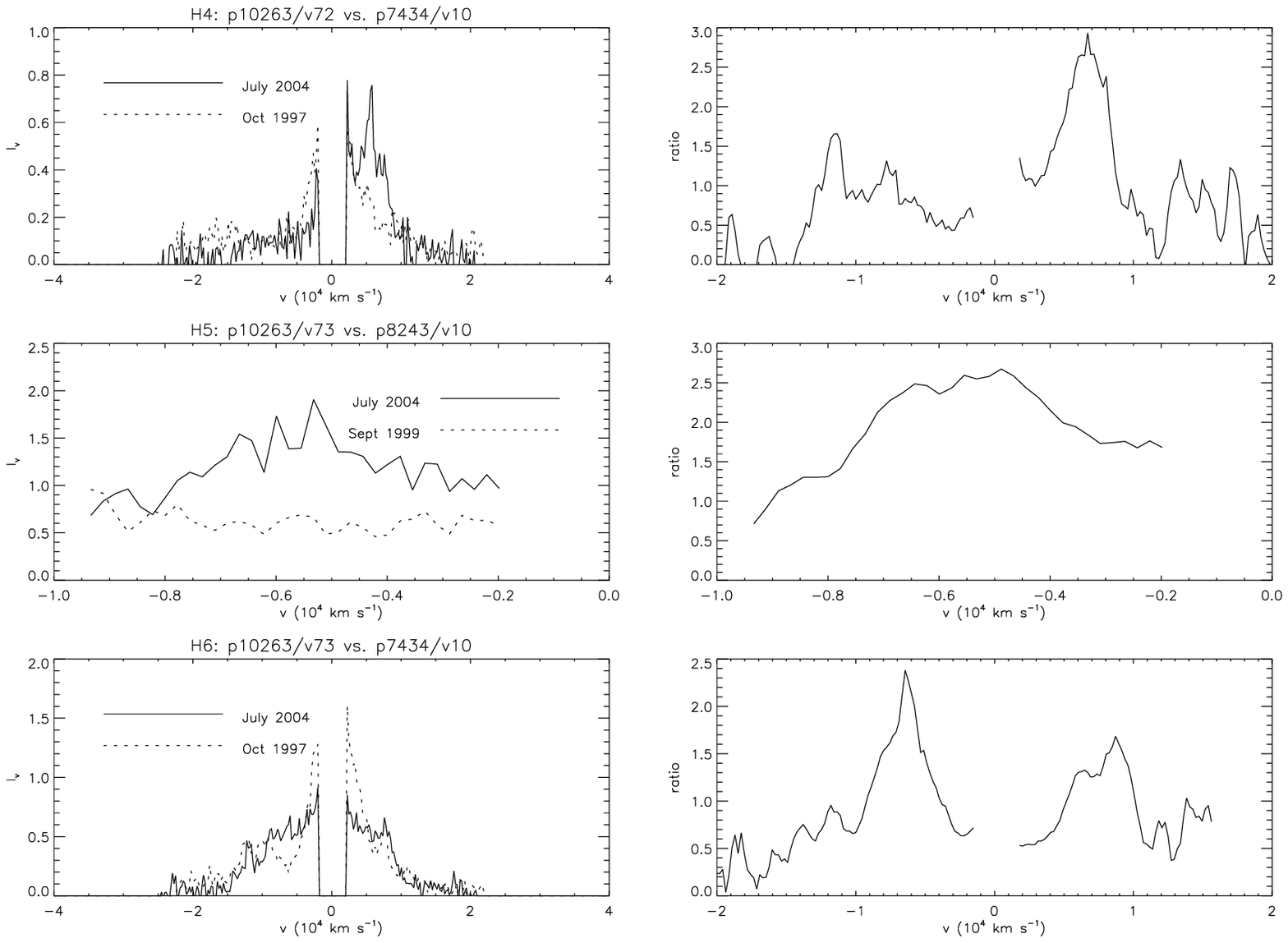}
\end{center}
\caption{Left: Average H$\alpha$ intensity, $I_v$, in units of $10^{-17}$ erg cm$^{-2}$ s$^{-1}$ (km s$^{-1}$)$^{-1}$ (arcsec)$^{-2}$, plotted 
vs. velocity ($v$).  Each panel is labeled according to the scheme in Fig.~\ref{fig:montage_h_alpha}.  In all of the panels, the solid 
curve corresponds to the observation listed first.  Right: Ratio of the new to the old intensities.}
\label{fig:local_h_alpha}
\end{figure}

\begin{figure}
\begin{center}
\includegraphics[width=5.5in]{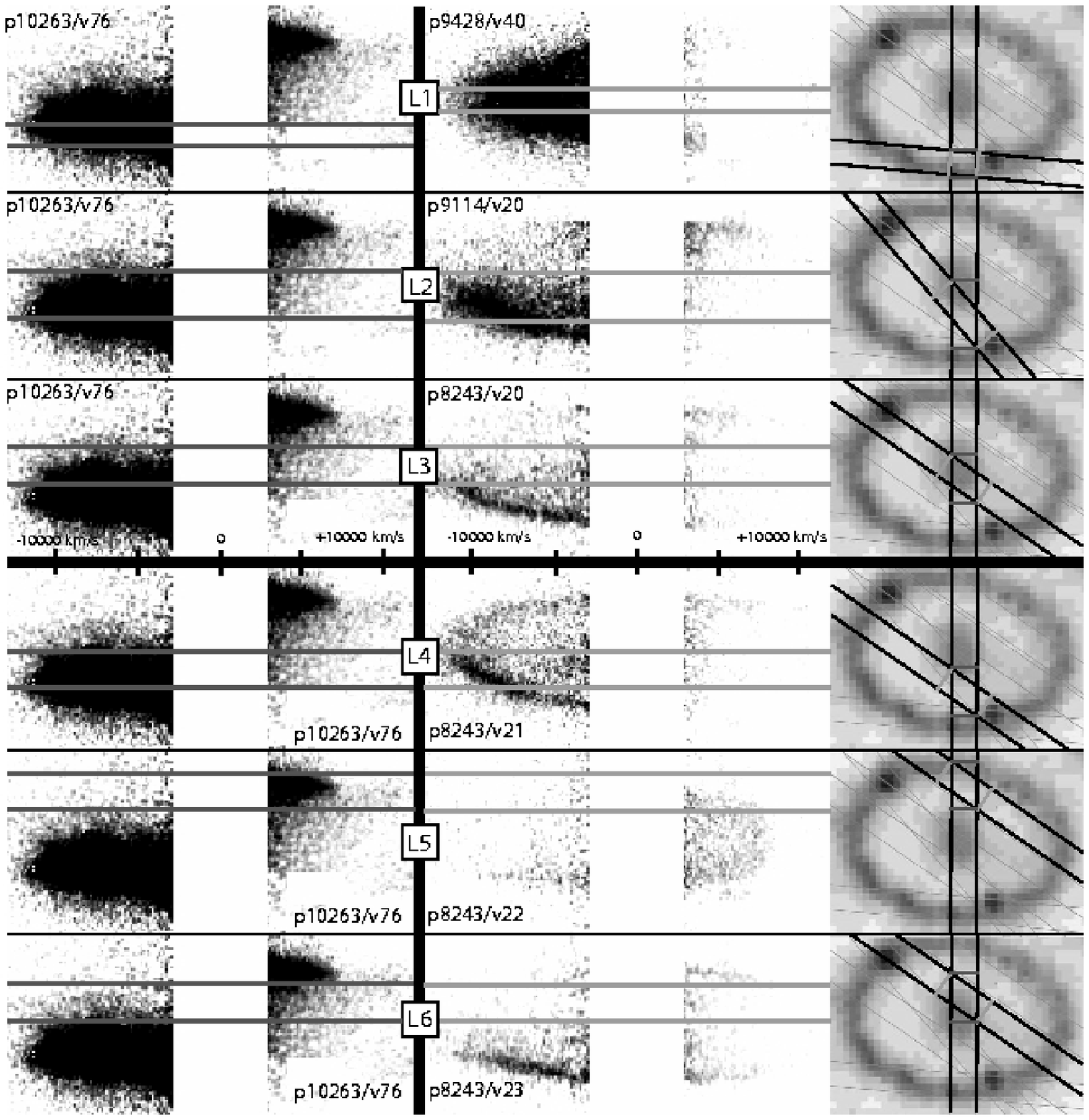}
\end{center}
\caption{Montage of the 0$\farcs$2 G140L observations used to track the local evolution of Ly$\alpha$ emission.  Each pair of 
observations is labeled as shown, with the newer (2004) observation being displayed in the left column; the corresponding slit 
overlays are in the right column.  In all of the slit overlay panels, North is down.  The newer (2004) observation always has its slit 
position aligned parallel to the North-South line.  The region of overlap is bracketed by the lines shown.  There are four points of overlap and summing between the innermost or outermost ones, in the vertical direction, has a negligible effect on the intensity ratios obtained (right panel of Fig.~\ref{fig:local_ly_alpha}).}
\label{fig:montage_ly_alpha}
\end{figure}

\begin{figure}
\begin{center}
\includegraphics[width=5in]{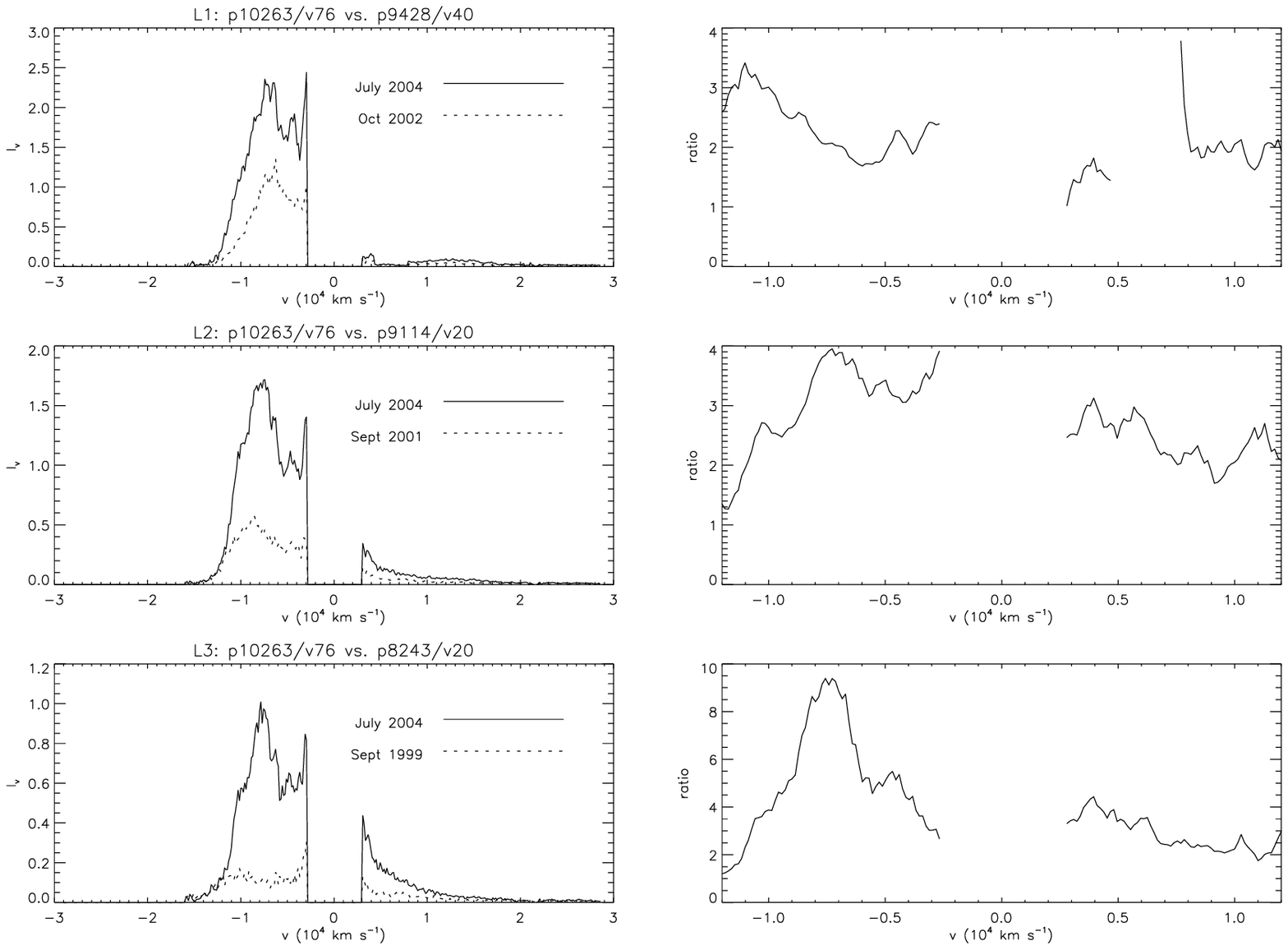}
\includegraphics[width=5in]{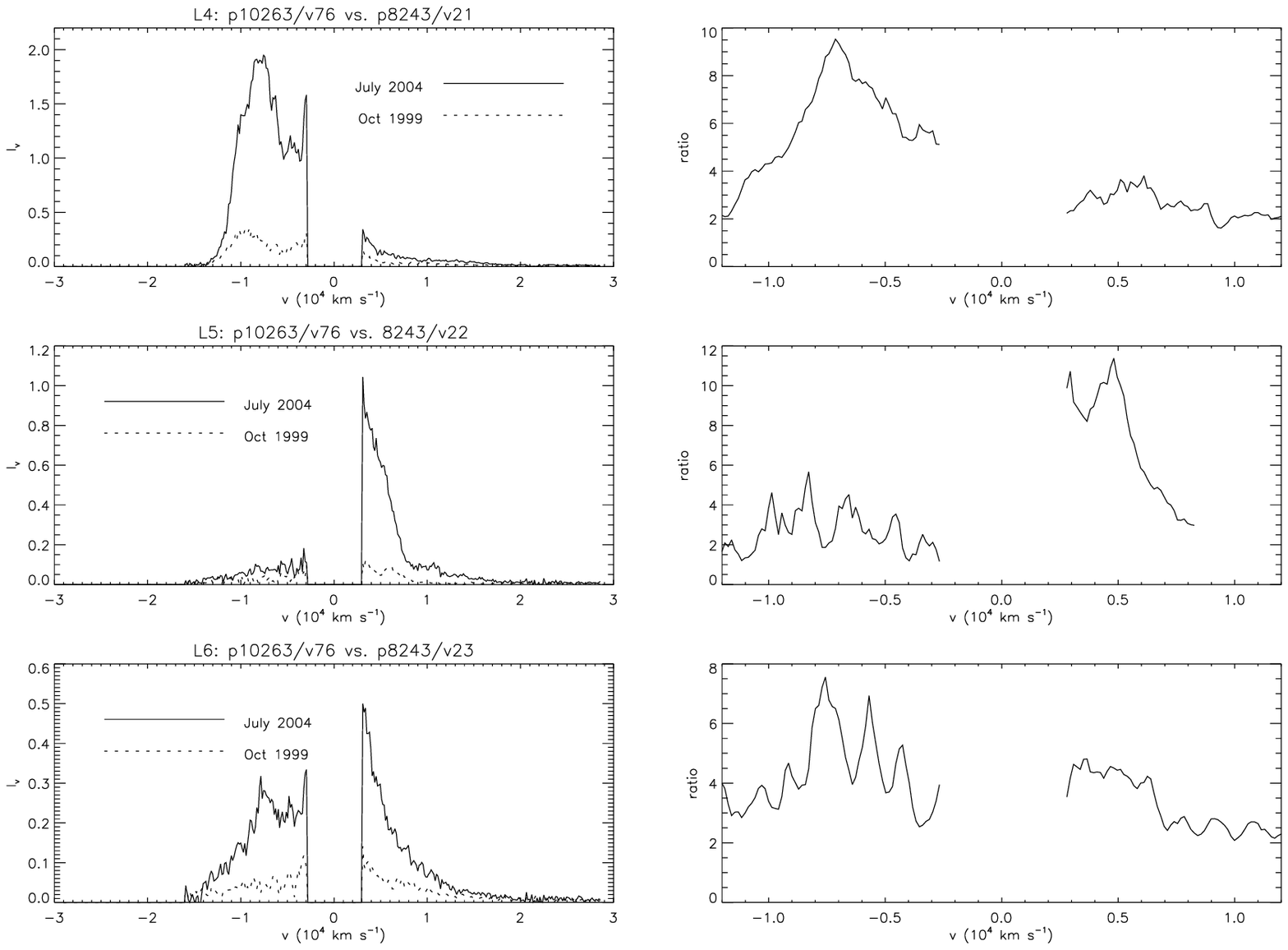}
\end{center}
\caption{Left: Average Ly$\alpha$ intensity, $I_v$, in units of $10^{-15}$ erg cm$^{-2}$ s$^{-1}$ (km s$^{-1}$)$^{-1}$ (arcsec)$^{-2}$, plotted 
vs. velocity ($v$).  Each panel is labeled according to the scheme in Fig.~\ref{fig:montage_ly_alpha}.  In all of the panels, the solid curve corresponds to the observation listed first.  Right: Ratio of the new to the old intensities.}
\label{fig:local_ly_alpha}
\end{figure}

Since no two subsequent observations with the 0$\farcs$2 slit have the same slit positions, we can only measure the evolution of intensity in the areas of overlap, which are in the form of parallelograms, labeled H1 to H6 (Fig.~\ref{fig:montage_h_alpha}) and L1 to L6 (Fig.~\ref{fig:montage_ly_alpha}), for G750L and G140L data, respectively. In each parallelogram, we are looking down a tube into the supernova debris, since there is a unique correspondence, given by equation (\ref{eq:doppler}), between the Doppler shift of the observed emission and the depth in the supernova debris.  To locate these parallelograms, we measure the differences in right ascension and declination from a given acquisition star and calculate the angular displacement between a pair of observations.  We then shift the observations to a common coordinate system and measure the parallelograms directly from the superimposed images.  For each slit orientation, we sum the pixels in the spatial direction (see Figs.~\ref{fig:montage_h_alpha} \& \ref{fig:montage_ly_alpha}) between the boundaries of the parallelogram and plot the average intensity (Figs.~\ref{fig:local_h_alpha} \& \ref{fig:local_ly_alpha}).  We find that summing between different points, in the spatial direction, has a negligible effect on the intensity ratios obtained, which are smoothed (using a width of 5 pixels for the smoothing window) to remove anomalies due to noise or insufficient signal.  Uncertainties in these and all subsequent measurements of reverse shock emission are dominated entirely by systematic errors due to contamination by emission from components other than the reverse shock.  Although it is difficult to judge the fractional contribution of these contaminants, we estimate that the uncertainties are typically $\sim\pm 10 \%$.

\subsection{H$\alpha$ EMISSION}

We first discuss the G750L observations.  Both blueshifted and redshifted emission features from the reverse shock are prominent in the more recent spectra on the left column of Fig.~\ref{fig:montage_h_alpha}, especially just inside the inner circumstellar ring.  But unfortunately, the regions where the reverse shock emission is brightest in these spectra usually do not overlap with the regions observed in the earlier spectra.  For example, consider the two spectra overlapping in parallelogram H1.  It is clear that the earlier (p8243/v10, 1999 September 18) spectrum is dominated by relatively low velocity ($\lesssim 3000$ km s$^{-1}$) H$\alpha$ and [O~{\sc i}] $\lambda\lambda$6300, 6364 emission from the supernova core.  Hence, the corresponding flux ratios are not representative of emission from the reverse shock.  The later (p10263/v71, 2004 July 18) spectrum is likewise dominated on the blue side by H$\alpha$ and [O~{\sc i}] core emission.  The very strong, blueshifted reverse shock emission that is evident on the north side of this spectrum does not fall in parallelogram H1. Redshifted emission in the velocity range $6500 \lesssim v \lesssim 8500$ km s$^{-1}$ on the southern edge of the later spectrum may be attributed to H$\alpha$ emission by the reverse shock and also by [S~{\sc ii}]; this emission is not evident in the earlier spectrum.  Likewise, very little reverse shock emission is present in parallelogram H2.

Parallelograms H3 and H4 largely overlap and both capture the evolution of the redshifted emission from the reverse shock on the southern side of the debris.  In H3, the average intensity increases from 1999 September 18 to 2004 July 18 by a maximum factor of 2.0, while in H4, it increases from 1997 October 6 to 2004 July 18 by a maximum factor of 2.9.

Parallelogram H6 captures the reverse shock only at high redshift ($v \gtrsim 8000 $ km s$^{-1}$) and high blueshift ($v \lesssim -8000 $ km s$^{-1}$).  In these velocity ranges, the average intensity increases from 1997 October 6 to 2004 July 23 by factors of 1.7 and 1.5, respectively.

Parallelogram H5 contains the only pair of H$\alpha$ observations which cleanly captures the surface emission from the near side of the debris.  The spectra shown in Fig.~\ref{fig:local_h_alpha} represent only the emission seen in the blueshifted streaks in the respective spectra (p8243/v10 and p10263/v73). The average intensity in these spectra increased by a maximum factor of 2.7 from 1999 September 18 to 2004 July 23.

In summary, in the regions where we can measure the evolution of H$\alpha$ emission from the reverse shock, both the blueshifted and redshifted average intensities have increased by maximum factors $\sim$ 3, over periods of about 5 and 7 years, respectively.

\subsection{Ly$\alpha$ EMISSION}

The increase in average intensity is more dramatic in the G140L observations.  The data, shown in Fig.~\ref{fig:montage_ly_alpha}, appear to be clean of contaminants and thus we believe that the intensity ratios plotted are reliable.  

Parallelogram L1 overlays the northern side of the circumstellar ring and captures blueshifted emission from the reverse shock on the near side of the debris. In this region, the maximum average intensity has increased by a factor of 3.4 from 2002 October 29 to 2004 July 20. 

Parallelograms L2 and L4 are partly coincident on the northern side of the debris and both capture much of the blueshifted reverse shock emission.  In L2, we see an increase by a factor of 3.9 from 2001 September 24 to 2004 July 20. In L4, we see an increase by a factor of 9.4 from 1999 October 7 to 2004 July 20.

Parallelograms L3, L5 and L6 give us a glimpse of the evolution of the redshifted Ly$\alpha$.  The reverse shock is most prominent in L5, where we see an increase by a factor of 5.7 from 1999 October 12 to 2004 July 20.  Parallelograms L3 and L6 show increases of 9.4 and 7.5, from 1999 Sept 28 and Oct 13, respectively, to 2004 July 20.

We thus see much greater increases of reverse shock emission in Ly$\alpha$ than we do in H$\alpha$.  For example, from September 1999 to July 2004, the blueshifted Ly$\alpha$ in L4 increased by a factor of 9.4, while the blueshifted H$\alpha$ in H5 increased by a factor of 2.7.  In the same period, the redshifted Ly$\alpha$ in L3 increased by a factor of 9.4, while the redshifted  H$\alpha$ in H3 increased by a factor of 2.0.  Unfortunately, however, the regions where we can clearly measure the evolution of the reverse shock emission in H$\alpha$ and Ly$\alpha$ do not overlap.  Therefore, we cannot be sure that the ratio of Ly$\alpha$ to H$\alpha$ emission from the reverse shock has increased.

The redshifted emission from the reverse shock seen in H$\alpha$ has a very different velocity profile from that seen in Ly$\alpha$ (compare Figs.~\ref{fig:montage_h_alpha} \& \ref{fig:montage_ly_alpha}).  As we discuss below, the Ly$\alpha$ emission seen from the reverse shock is profoundly modified by resonant scattering, and so we cannot meaningfully compare the evolution of the emission in H$\alpha$ and Ly$\alpha$ from this part of the reverse shock.    

\section{H$\alpha$ VS. Ly$\alpha$ EMISSION}
\label{sect:compare}

\begin{figure}
\begin{center}
\includegraphics[width=5in]{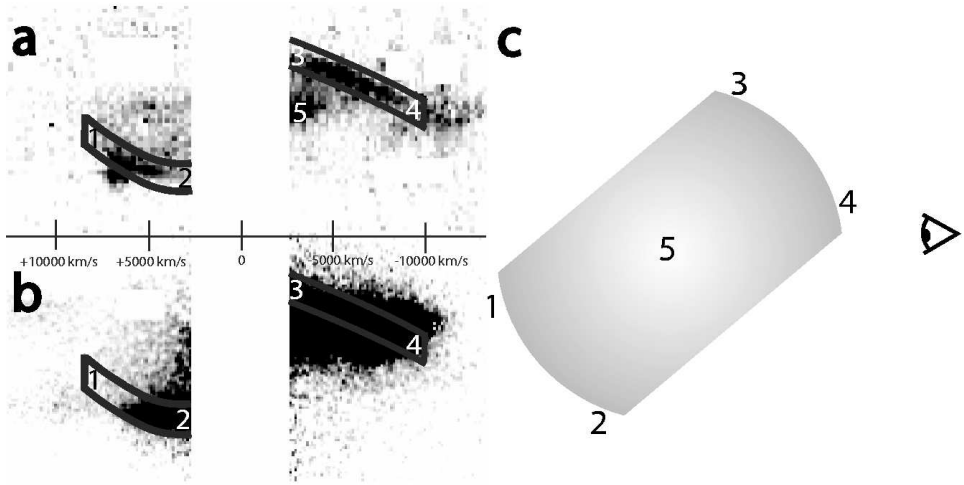}
\end{center}
\caption{(a) H$\alpha$ surface emission (p10263/v71) from the reverse shock isolated by masks.  (b) Ly$\alpha$ surface emission (p10263/v76) with the 
same masks applied.  (c) Schematic representation of the supernova debris with the boundary being defined by the reverse shock.  For 
freely-expanding debris, there is a unique correspondence between velocity and the origin of the emission along the line of sight.}
\label{fig:geometry}
\end{figure}

\begin{figure}
\begin{center}
\includegraphics[width=5in]{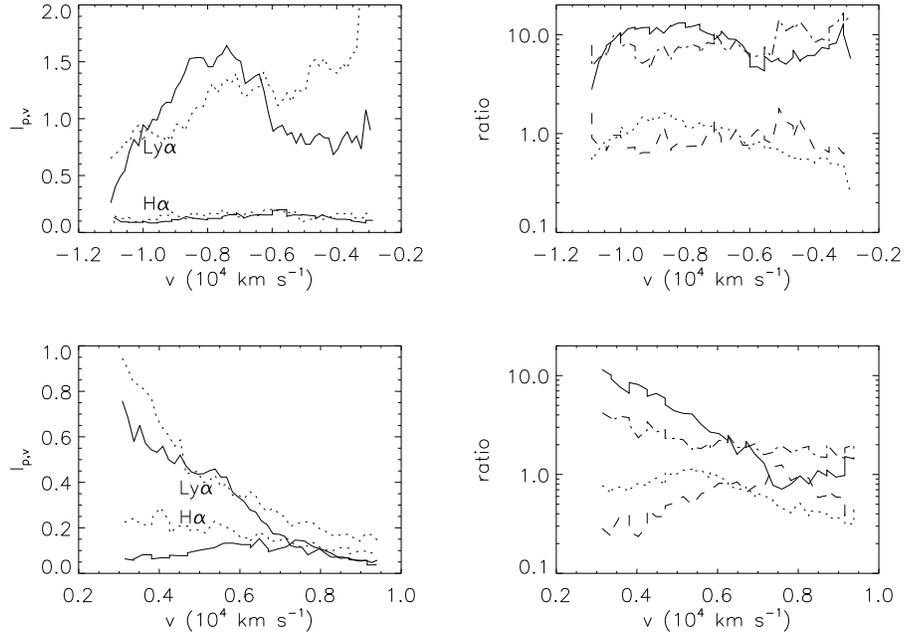}
\end{center}
\caption{Top left: Surface (solid curve) vs. interior (dotted) average intensity, $I_{p,v}$, in units of 10$^6$ photons 
cm$^{-2}$ s$^{-1}$ (km s$^{-1}$)$^{-1}$ (arcsec)$^{-2}$, for both Ly$\alpha$ and H$\alpha$ emission from the near side of the debris.  Top right: 
Ratio of Ly$\alpha$-to-H$\alpha$ surface emission (solid), Ly$\alpha$-to-H$\alpha$ interior emission (dot-dashed), surface-to-interior 
emission for Ly$\alpha$ (dotted) and H$\alpha$ (dashed), for emission from the near side of the debris.  Bottom left \& right: Same as for top left and right respectively, but for emission from the far side of the debris.}
\label{fig:compare}
\end{figure}

Our most recent (July 2004) STIS observations give us our first opportunity to compare the surface and interior emission in H$\alpha$ (p10263/v71) and 
Ly$\alpha$ (p10263/v76) through the same slit location and at the same epoch.  Since the H$\alpha$ photons are unaffected by resonant scattering (M03), they are a more reliable tracer of the reverse shock intensity than the Ly$\alpha$ ones.  Furthermore, as discussed in \S\ref{sect:cleaning}, we can differentiate the surface from the interior emission more clearly in H$\alpha$ than Ly$\alpha$.  We construct masks to isolate the reverse shock, using the G750L observation as a template (Fig.~\ref{fig:geometry}a).  We spatially sum the data and plot the average intensities in Fig.~\ref{fig:compare}.  Positions 1 to 5 are labeled on Fig.~\ref{fig:geometry}c; their corresponding positions on the G750L and G140L data are shown in Figs.~\ref{fig:geometry}a \& \ref{fig:geometry}b, respectively (assuming that all the emission comes from freely-streaming hydrogen atoms).

Figure \ref{fig:compare} shows that the ratio of Ly$\alpha$ to H$\alpha$ surface emission peaks at 13.1 on the near side and dips to a minimum of 2.8.  From the H$\alpha$-to-Ly$\alpha$ production ratio of 0.21 alone (M03), we expect 4.8 Ly$\alpha$ photons to be emitted for each H$\alpha$ photon.  The fact that we observe more than twice this ratio is evidence of resonant scattering.  Half of the Ly$\alpha$ photons produced on the near side of the debris are emitted outward towards the observer.  The other half are emitted inward towards the unshocked debris.  Most of these are backscattered out of the debris, towards the observer.  H$\alpha$ photons are unaffected by resonant scattering.  Hence, on the near side of the debris, we expect to observe a Ly$\alpha$/H$\alpha$ ratio that is about twice the production ratio - i.e., a factor $\sim$ 10.

On the far side of the debris, the Ly$\alpha$-to-H$\alpha$ ratio ranges from 0.71 to 11.5.  We expect to observe a ratio less than the production ratio because most of the Ly$\alpha$ photons produced there are scattered outwards, away from the observer.  This is the case for Ly$\alpha$ photons coming from the region labeled ``1'' in Fig.~\ref{fig:geometry}c.  However, we do see Ly$\alpha$ photons coming from region 2.  Evidently, these photons are produced near the limb of the debris and can reach the observer without having to propagate through the hydrogen atoms in the supernova debris.

\section{INTERIOR EMISSION}
\label{sect:interior}

Besides comparing surface emission in H$\alpha$ and Ly$\alpha$, Fig.~\ref{fig:compare} also shows the ratio of the surface to interior emission for both H$\alpha$ (p10263/v71) and Ly$\alpha$ (p10263/v76).  We have masked out the core emission in H$\alpha$, which originates from Location 5.  (Ly$\alpha$ emission cannot emerge from the supernova core, owing to the high optical depth of the debris.)  For Ly$\alpha$, the ratio of surface to interior emission peaks at 1.6 on the near side and 1.1 on the far side.  For H$\alpha$, the ratios are 1.8 and 1.2 respectively.  These ratios show that interior emission is comparable to surface emission in both cases, which means that the ratio of Ly$\alpha$ to H$\alpha$ interior emission tracks its corresponding surface ratio closely.  Resonant scattering can account for the surface emission ratios discussed previously, but not for the origin of the interior emission (see \S\ref{sect:discussion}).  Overall, the ratio of the total H$\alpha$ to Ly$\alpha$ flux, which includes both surface and interior emission, is 0.17, a value consistent with the production ratio (0.21).

\begin{figure}
\begin{center}
\includegraphics[width=6.5in]{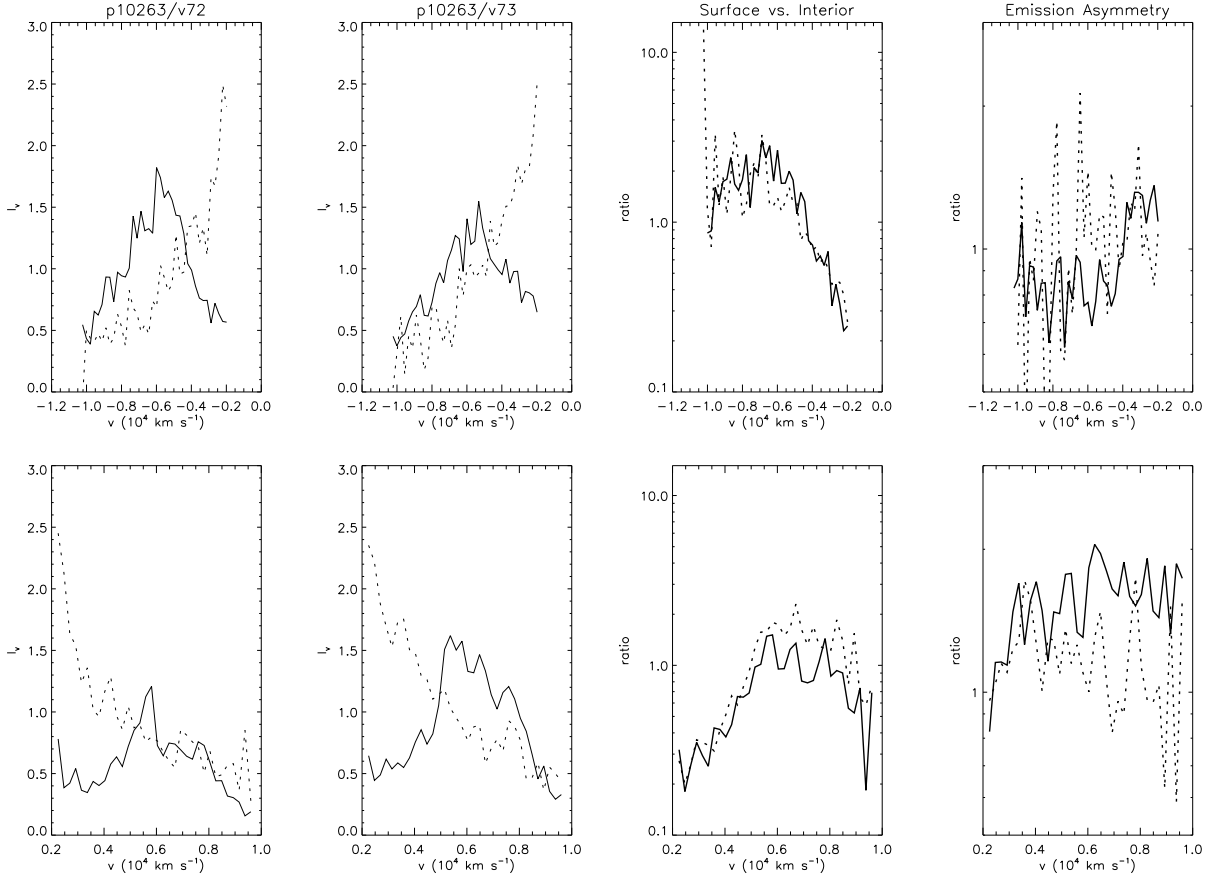}
\end{center}
\caption{Top 1st: Surface (solid curve) vs. interior (dotted) average intensity, $I_v$, in units of $10^{-17}$ erg cm$^{-2}$ s$^{-1}$ (km s$^{-1}$)$^{-1}$ (arcsec)$^{-2}$, plotted vs. velocity ($v$), for blueshifted emission from observation p10263/v72.  Top 2nd: Same as top 1st, but for p10263/v73.  Top 3rd: Ratio of surface-to-interior emission for p10263/v72 (bold solid curve) and p10263/v73 (bold dotted curve).  Top 4th: Ratio of surface-to-surface (bold solid curve) and interior-to-interior (bold dotted curve) emission for p10263/v73 vs. p10263/v72.  These ratios describe the east-to-west asymmetry in the emission.  Bottom: Same as for the top row, but for redshifted emission.}
\label{fig:interior}
\end{figure}

We measure the H$\alpha$ surface and interior emission for observations p10263/v72 and p10263/v73 as well.  These observations have the same slit orientations as p10263/v71, but are positioned just west and east of it respectively (Fig.~\ref{fig:overlays}a); p10263/71 is centered on the remnant.  We find that, as in the case of p10263/v71, the interior emission tracks its surface counterpart closely (Fig~\ref{fig:interior}).  On the northern (i.e., near) side of the remnant, the surface emission seen in the western slit is generally brighter than that in the east.  But on the southern (i.e., far) side of the remnant, the emission seen in the eastern slit is stronger than that in the west.  These results imply a departure of the reverse shock surface emission from cylindrical symmetry.   We note that the X-ray (Burrows et al. 2000) and radio (Gaensler et al. 1997) images also show departures from cylindrical symmetry.  No obvious asymmetry is evident in the interior emission.

\section{GLOBAL EVOLUTION OF REVERSE SHOCK EMISSION}
\label{sect:Halphalightcurve}

To construct a global light curve for the reverse shock emission in H$\alpha$, we first obtain profiles of the total velocity-dependent H$\alpha$ flux using a combination of the 0$\farcs$2, 0$\farcs$5 and 2$^{\prime\prime}$ observations and the data reduction procedures described in \S\ref{sect:reduction}.  Again, the statistical errors are negligible and the main source of uncertainty comes from the systematic error, which stems from not knowing the shape of the emission line profile at low velocities.  Michael et al. (2002) demonstrated, albeit in a slightly different context (i.e., in the X-ray), that the line profile is strongly dependent on the geometry of the system.  For example, if the system is a cylindrical ring expanding in the equatorial plane, a ``double-horned'' profile results, with very little low-velocity emission and the ``horns'' centered at $\pm v_s$, where $v_s$ is the freely-streaming, projected velocity of the debris in the rest frame of the reverse shock.  By contrast, a radially-expanding spherical shell of gas produces a square-topped profile and a filled, uniform sphere produces a parabolic profile.

\begin{figure}
\begin{center}
\includegraphics[width=4.5in]{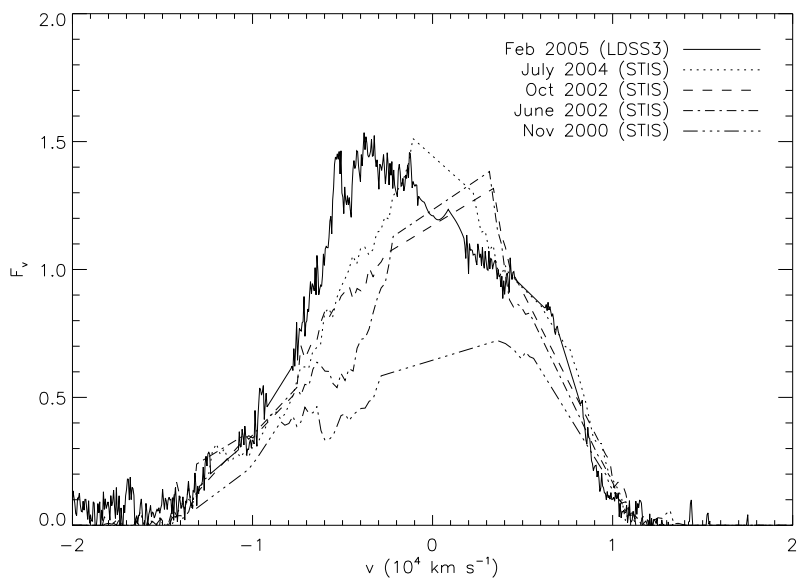}
\end{center}
\caption{Total H$\alpha$ velocity-dependent flux, $F_v$, in units of $10^{-13}$ erg cm$^{-2}$ s$^{-1}$ (km s$^{-1}$)$^{-1}$, from the reverse shock.  The Magellan/LDSS3 profile shown is from Smith et al. (2005).}
\label{fig:profiles}
\end{figure}

\begin{figure}
\begin{center}
\includegraphics[width=4.5in]{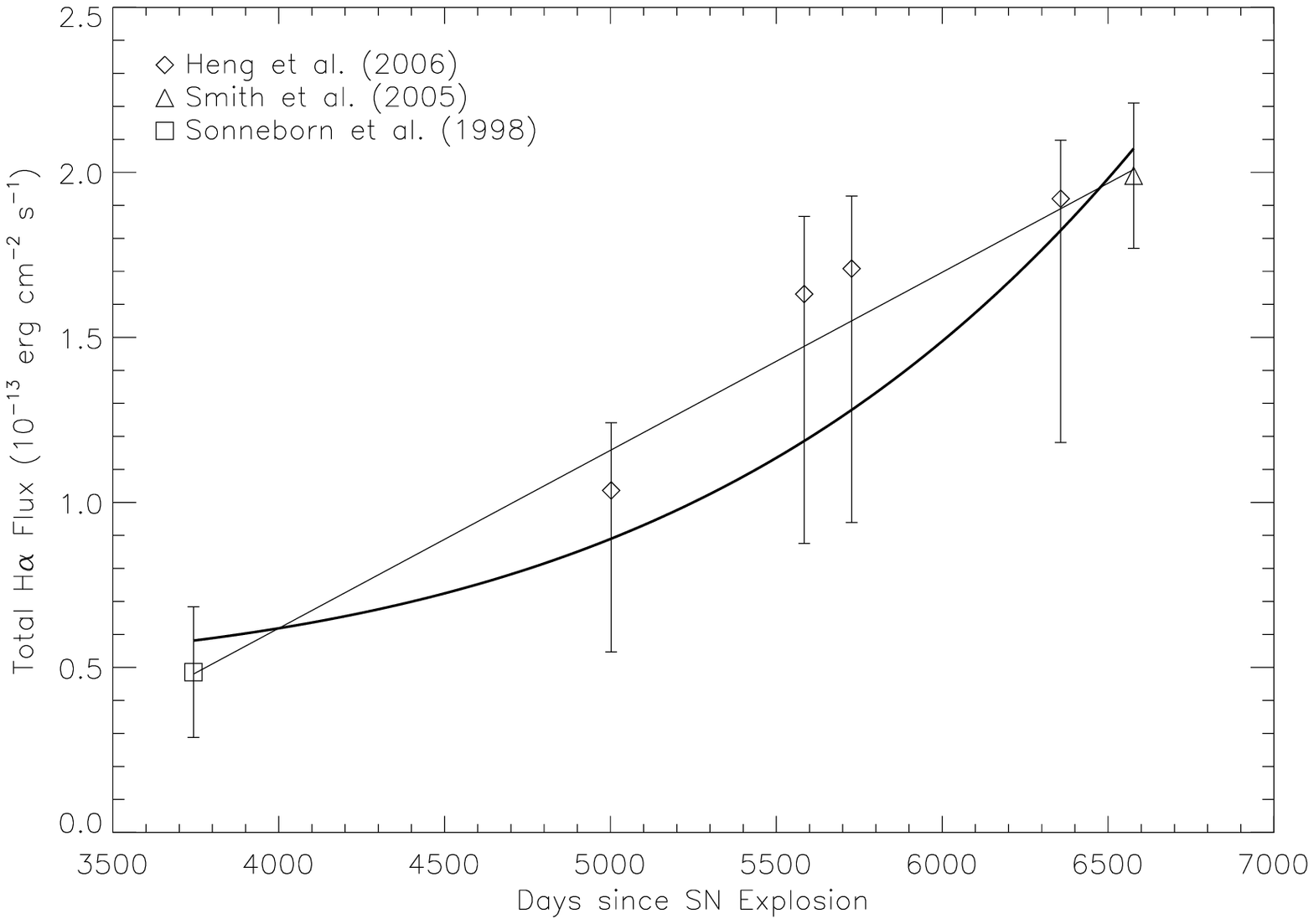}
\end{center}
\caption{Global evolution of H$\alpha$ flux from the reverse shock.  The thin line and bold curve correspond to $F \propto t$ and $F \propto t^5$ fits, respectively.}
\label{fig:lightcurve}
\end{figure}

To estimate the total line emission, we interpolate linearly over the masked data to obtain the H$\alpha$ profiles as indicated in Fig.~\ref{fig:profiles}, and integrate over velocity to obtain the total fluxes listed in Table~\ref{tab:fluxes} and plotted in Fig.~\ref{fig:lightcurve}.  We include the profile obtained by S05.  We estimate very conservative lower bounds (typically $\sim 45 \%$) of the fluxes by excluding the interpolated profiles in the velocity ranges  $-3000 \lesssim v \lesssim 3000$ km s$^{-1}$ covered by the central slit.  We estimate upper bounds (typically $\sim 15 \%$) by making spline fits to interpolate the overall profiles through the central portions. 

Since the 0$\farcs$8 slit used in the Magellan observation by S05 did not fully capture the eastern and western portions of SNR 1987A, the resulting profile underestimates the amount of low-velocity emission actually present.  Hence, it is not surprising that portions of our profiles in Fig.~\ref{fig:profiles} exceed the Magellan/LDSS3 one.  The minimum and maximum velocities to which the flux extends appear to be unchanged, and they suggest that the expanding debris is more extended towards than away from us.  We are unable to determine these to a greater accuracy, as the high-velocity emission is contaminated by [O~{\sc i}] and [S~{\sc ii}] emission.  Using 2$^{\prime\prime}$ G750L data from STIS, S98 measured a total Ly$\alpha$ flux of ($1.25 \pm 0.51) \times 10^{-12}$ erg cm$^{-2}$ s$^{-1}$ on Day 3743.  Since the production ratio of H$\alpha$ to Ly$\alpha$ photons is 0.21, we infer an approximate H$\alpha$ flux of ($4.86 \pm 1.98) \times$10$^{-14}$ erg cm$^{-2}$ s$^{-1}$.  This figure is uncertain because the presence of resonant scattering results in an unknown conversion factor between the Ly$\alpha$ photons produced and those actually observed.  Nevertheless, we include this data point in our H$\alpha$ light curve (Fig.~\ref{fig:lightcurve}), along with the ($1.99 \pm 0.22) \times$10$^{-13}$ erg cm$^{-2}$ s$^{-1}$ result obtained by S05 on Day 6577.  We then find that the total reverse shock flux has increased by a factor of about 4.1 over 2834 days ($\sim$ 8 years).  

There are two limiting cases of how the reverse shock can behave.  The self-similar solution of Chevalier (1982) shows that the radius of the reverse shock varies as $R \propto t^\frac{n-3}{n-s}$, from which we infer that the flux across the shock is
\begin{equation}
F \propto \rho v R^2 \propto t^\frac{2n-9+4s-ns}{n-s},
\end{equation}
where the outer part of the freely-expanding supernova debris ($v=R/t$) and the ambient medium have the density profiles $\rho \propto t^{-3} v^{-n}$ and $\rho_m \propto R^{-s}$ respectively.  For a uniform medium ($s=0$) and $n=9$ (Eastman \& Kirshner 1989), $F \propto t$.  However, the blast wave is now overtaking density protrusions on the equatorial ring and each encounter will send a reflected shock inward towards the reverse shock.  If the merging of these shocks brings the reverse shock to a halt, then we have $F \propto t^{n-4} = t^5$.  Fitting a linear function, $F = F_0 (t-t_0)$, to the data, we obtain $F_0=5.40\times$10$^{-17}$ erg cm$^{-2}$ s$^{-1}$ day$^{-1}$ and $t_0=2853$ days.  If instead we fit the power law, $F = F_1 t^5 + F_2$, then we have $F_1=1.29\times 10^{-32}$ erg cm$^{-2}$ s$^{-1}$ day$^{-5}$ and $F_2=4.87\times 10^{-14}$ erg cm$^{-2}$ s$^{-1}$.  Note that the slope of this line is determined almost entirely by the initial STIS observation of Ly$\alpha$ and by the final Magellan observation.  The intervening STIS observations of H$\alpha$ are consistent with this fit, but their uncertainties are so great that they do not significantly constrain it, and we are unable to discern between the two limiting cases discussed.

\section{DISCUSSION}
\label{sect:discussion}

Our analysis of the STIS data presented above raises more questions than it answers, and a detailed explanation of the observations will require theoretical modeling beyond the scope of the present paper.  In this section, we summarize the main puzzles and suggest possible solutions.

\subsection{THE Ly$\alpha$/H$\alpha$ RATIO}
\label{sect:resonant}

As shown in \S\ref{sect:compare}, resonant scattering strongly suppresses the propagation of Ly$\alpha$ through the debris of SNR 1987A.  Photons come into resonance when the ``blueness" of their frequency is counteracted by the expansion redshift of the debris.  A photon emitted in the blue wing of the emission profile sees an increasing optical depth and has a very high probability of being scattered before it penetrates far into the debris.  By contrast, a photon emitted in the red wing becomes more redshifted as it travels through the debris, and has a good chance of passing through it.  M03 calculated that the Sobolev optical depth for resonant scattering of a Ly$\alpha$ photon is $\tau_s \approx 1000$.  For photons emitted on the near side of the debris, half make the journey inward towards the supernova core.  Most of these are backscattered towards the observer, effectively doubling the number of expected Ly$\alpha$ from the near side.  On the far side, the very same effect suppresses the number of Ly$\alpha$ photons which make the journey through the debris and to the observer.

M03 investigated this process using one-dimensional, ``two-stream'' (Schuster 1905), Monte Carlo simulations, but remarked that a more detailed analysis was beyond the scope of their study.  Such an analysis will have to model the resonant scattering of Ly$\alpha$ emission, in three dimensions, through an expanding debris, of characteristic thickness $L \sim vt \sim 10^{18}$ cm, where $v \sim 10,000$ km s$^{-1}$ and $t=18$ years.  If the path length traveled by the photon in a single scattering is $l$, then it will be redshifted by a amount $\nu_0 l / ct$ due to the expansion.  The path length is in turn computed from the optical depth, $\tau$, generated from the probability for scattering, $e^{-\tau}$.  A complication arises from the fact that there is a correlation between $l$, the frequency of the photon, the temperature of the gas in the debris and the projected velocity of the atom it encounters (e.g., Zheng \& Miralda-Escud\'{e} 2002).  If the Doppler core is involved, the surface of last scattering is located at a distance $\sim 10^{-4} T_{100}^{1/2} L$ from the emission surface, where $T_{100}=T/100$ K and $T$ is the temperature of the gas.  For values of parameters representative of the freely-expanding debris of SNR 1987A ($n \approx 100$ cm$^{-3}$, $T \approx 100$ K), we find that about 99\% of the Ly$\alpha$ photons are resonantly backscattered, resulting in a Ly$\alpha$-to-H$\alpha$ ratio $\sim 10$ on the near side of the debris.  The details of these calculations are somewhat complicated and will be presented elsewhere (Heng \& McCray, in preparation).

\subsection{IS CHARGE TRANSFER RESPONSIBLE FOR THE ``INTERIOR'' EMISSION?}
\label{sect:interior2}

We have shown in \S\ref{sect:interior} that ``interior'' emission is present in both H$\alpha$ and Ly$\alpha$ and that it is comparable in strength to the surface emission.  If we stick to the assumption that the photons are emitted by freely-streaming hydrogen atoms, then we need to identify an emission mechanism which produces photons at a location intermediate between the reverse shock surface and the supernova core, with a production ratio of Ly$\alpha$/H$\alpha \approx 5$.  The interior emission, as shown in Fig.~\ref{fig:compare}, is present across the broad velocity range $\sim \pm 10,000$km s$^{-1}$.

One possible mechanism is the interaction of the atoms with relativistic particles that have been accelerated at the reverse shock and have diffused upstream into the unshocked supernova debris.  We cannot say with confidence whether this mechanism can explain the observed interior emission, since we have no quantitative theory to estimate the flux of relativistic particles in the reverse shock.  

Another possible explanation for the interior emission is that it does not actually originate from freely-streaming hydrogen atoms in the supernova debris, but instead comes from hydrogen atoms resulting from charge transfer/exchange reactions of the freely-streaming atoms with protons in the hot gas beyond the reverse shock.  The resulting atoms will have a broad distribution of radial velocities similar to that of the shocked protons, centered at $\sim 4000$ km s$^{-1}$ rather than $\sim 10000$ km s$^{-1}$.  Photons emitted by subsequent excitation of these atoms will appear to originate from the interior of the supernova debris. However, we cannot be sure that this mechanism can account quantitatively for the observed emission without a detailed calculation, which is beyond the scope of the present paper.   

\subsection{THE LIGHT CURVE OF THE REVERSE SHOCK EMISSION?}

The global evolution of the H$\alpha$ emission (Fig.~\ref{fig:lightcurve}) is of great interest, since one can directly infer from it the net flux of atoms across the reverse shock.  As S05 have pointed out, the ionizing radiation originating from beyond the reverse shock surface may photoionize the hydrogen atoms before they reach it and suppress the reverse shock emission.  At present, the ionizing flux is increasing more rapidly than the reverse shock emission.  If present trends continue, the reverse shock emission will vanish in $\lesssim 7$ years from now.  

However, as we have shown, the STIS observations do not significantly constrain this evolution.  As described by S05, it is possible to track the global evolution of the reverse shock emission in H$\alpha$ with ground-based observations.  To do this accurately, we must obtain deep, low- or moderate-resolution spectra using a slit wide enough to include the entire reverse shock surface.  This avoids the technical problems discussed in \S\ref{sect:Halphalightcurve}.  

If the Cosmic Origin Spectrograph (COS) is installed on the Hubble Space Telescope, it will be possible to measure both the local and global emission of Ly$\alpha$ from the reverse shock.  Such observations will be extremely valuable for understanding the effects of resonant scattering and charge transfer as well.

\section{SUMMARY}
\label{sect:summary}

Using STIS data, we have tracked both the local and global evolution of the H$\alpha$ and Ly$\alpha$ emission from the reverse shock in SNR 1987A. The main results of our study are: 

1. The average H$\alpha$ intensity has increased locally by a factor $\sim$ 3 for both blueshifted and redshifted emission, over periods of about 5 and 7 years, respectively.  For the average Ly$\alpha$ intensity, the factors are both $\sim$ 9 over about 5 years.  However, we cannot compare these factors directly, as the regions where we can clearly measure the evolution of the reverse shock emission do not overlap.

2. In a comparison of the emission from the reverse shock in H$\alpha$ and Ly$\alpha$, taken through the same slit at the same time (July 2004), we find clear evidence that the transfer of Ly$\alpha$ radiation through the supernova debris is suppressed by resonant scattering.

3. In addition to the emission from the surface of the reverse shock, we see emission of comparable intensity in both H$\alpha$ and Ly$\alpha$ that appears to come from interior to the surface.  This emission may be the result of charge transfer reactions near the reverse shock surface.

4. In H$\alpha$ observations taken with three adjacent slit positions during July 2004, we find departures from cylindrical symmetry in the reverse shock surface emission, with the brighter emission towards the northwest and the southeast.

5. Using data from STIS, we construct the light curve of the reverse shock emission in H$\alpha$, which has brightened by a factor $\sim 4$ over about 8 years.  The STIS data are consistent with current ground-based observations by Magellan (S05), but have much greater systematic uncertainties.  Consequently, we are unable yet to determine the acceleration rate of the brightening of the reverse shock.  With future ground-based observations of H$\alpha$, we should be able to measure this light curve with sufficient accuracy to determine whether and when photoionization will turn off the reverse shock emission.

\acknowledgments \scriptsize

K.H. wishes to thank Dimitri Veras, Eli Michael, James Aguirre, Joshua Schroeder, Brian Morsony and Jared Workman for many fruitful discussions and helpful suggestions regarding various technical issues.  Gratitude is shown to Peter Ruprecht and James McKown for outstanding technical support at JILA.

K.H. and S.Z. were supported by NASA through {\it Chandra} award G04-5072A to the University of Colorado.  C.S.J.P. acknowledges support from the Research Grant Council of Hong Kong.

\normalsize


\begin{table}
\begin{center}
\caption{STIS Observations Used}
\label{tab:observations}
\begin{tabular}{lrrrrr}
\tableline\tableline
\multicolumn{1}{c}{Label} & \multicolumn{1}{c}{Date} &
\multicolumn{1}{c}{Type} & \multicolumn{1}{c}{Slit Width} & 
\multicolumn{1}{c}{Slit P.A. (deg)} & \multicolumn{1}{c}{Exposure Time (s)} \\
\tableline
p10263/v73 & 2004 Jul 23 & G750L & 0$\farcs$2 & 179.7 & 5468 \\
p10263/v72 & 2004 Jul 18 & G750L & 0$\farcs$2 & 179.7 & 5468 \\
p10263/v71 & 2004 Jul 18 & G750L & 0$\farcs$2 & 179.7 & 5468 \\
p8243/v10 & 1999 Sept 18 & G750L & 0$\farcs$2 & -149.0 & 9983 \\
p7434/v10 & 1997 Oct 6 & G750L & 0$\farcs$2 & -139.0 & 9590 \\
p10263/v76 & 2004 Jul 20 & G140L & 0$\farcs$2 & 179.7 & 5350 \\
p9428/v40 & 2002 Oct 29 & G140L & 0$\farcs$2 & -95.3 & 10600 \\
p9114/v20 & 2001 Sept 24 & G140L & 0$\farcs$2 & -139.3 & 10700 \\
p8243/v23 & 1999 Oct 13 & G140L & 0$\farcs$2 & -124.8 & 10665 \\
p8243/v22 & 1999 Oct 12 & G140L & 0$\farcs$2 & -124.8 & 10665 \\
p8243/v21 & 1999 Oct 7 & G140L & 0$\farcs$2 & -124.8 & 10478 \\
p8243/v20 & 1999 Sept 28 & G140L & 0$\farcs$2 & -124.8 & 10478 \\
p10263/v74 & 2004 Jul 23 & G750L & 0$\farcs$5 & 179.7 & 5282 \\
p10263/v75 & 2004 Jul 23 & G750L & 0$\farcs$5 & 179.7 & 5282 \\
p9428/v35 & 2002 Oct 29 & G750L & 2$^{\prime\prime}$ & -93.8 & 4832 \\
p9328/v2 & 2002 June 8 & G750L & 2$^{\prime\prime}$ & 110.7 & 4970 \\
p8872/v2 & 2000 Nov 3 & G750L & 2$^{\prime\prime}$ & -92.8 & 4830 \\
\tableline
\end{tabular}
\end{center}
\end{table}

\begin{table}
\begin{center}
\caption{Total H$\alpha$ Flux Obtained from STIS Observations}
\label{tab:fluxes}
\begin{tabular}{lcc}
\tableline\tableline
\multicolumn{1}{c}{Date} & \multicolumn{1}{c}{Observations} &
\multicolumn{1}{c}{Flux} \\
\multicolumn{1}{c}{} & \multicolumn{1}{c}{} & \multicolumn{1}{c}{($10^{-13}$ erg cm$^{-2}$ s$^{-1}$)}\\
\tableline
Jul 2004 & p10263/v71+72+73+74+75 & 1.92$^{+0.18}_{-0.74}$\\
Oct 2002 & p9428/v35 & 1.73$^{+0.26}_{-0.78}$ \\
Jun 2002 & p9328/v2 & 1.63$^{+0.24}_{-0.76}$ \\
Nov 2000 & p8872/v2 & 1.04$^{+0.21}_{-0.49}$ \\
\tableline
\end{tabular}
\end{center}
\end{table}

\begin{references}

\reference{} Borkowski, K.J., Blondin, J., \& McCray, R.\ 1997, ApJ, 476, L31

\reference{} Burrows, C.J., et al. \ 1995, ApJ, 452, 680

\reference{} Burrows, D.N., et al. \ 2000, ApJ, 543, L149

\reference{} Cassatella, A., Fransson, C., van Santvoort, J., Gry, C., Talavera, A., Wamsteker, W., \& Panagia, N. \ 1987, A\&A, 177, L29

\reference{} Chevalier, R.A., \& Raymond, J.C. \ 1978, ApJ, 225, L27

\reference{} Chevalier, R.A.\ 1982, ApJ, 258, 790

\reference{} Chevalier, R.A., \& Dwarkadas, V.V. \ 1995, ApJ, 452, L45

\reference{} Crotts, A.P.S., Kunkel, W.E., \& McCarthy, P.J. \ 1989, ApJ, 347, L61

\reference{} Crotts, A.P.S., \& Kunkel, W.E. \ 1991, ApJ, 366, L73

\reference{} Crotts, A.P.S., Kunkel, W.E., \& Heathcote, S.R. \ 1995, ApJ, 438, 724

\reference{} Crotts, A.P.S., \& Heathcote, S.R. \ 2000, ApJ, 528, 426

\reference{} Eastman, R.G., \& Kirshner, R.P.\ 1989, ApJ, 347, 771

\reference{} Fransson, C., Grewing, M., Cassatella, A., Panagia, N., \& Wamsteker, W. \ 1987, A\&A, 177, L33

\reference{} Fransson, C., \& Lundqvist, P. \ 1989, ApJ, 341, L59

\reference{} Gaensler, B.M., Manchester, R.N., Staveley-Smith, L., Tzioumis, K., Reynolds, J.E., \& Kesteven, M.J. \ 1997, ApJ, 479, 845

\reference{} Hill, R.S., Landsman, W.B., Lindler, D., \& Shaw, R. \ 1997, HST Calibration Workshop, ed. S. Casertano, et al.

\reference{} Lawrence, S.S., Sugerman, B.E.K., Bouchet, P., Crotts, A.P.S., Uglesich, R., \& Heathcote, S. \ 2000, ApJ, 537, L123

\reference{} Luo, D., \& McCray, R. \ 1991, ApJ, 379, 659

\reference{} Luo, D., McCray, R., \& Slavin, J. 1994, ApJ, 430, 264

\reference{} Manchester, R.N., Gaensler, B.M., Wheaton, V.C., Staveley-Smith, L., Tzioumis, A.K., Bizunok, N.S., Kesteven, M.J., \& Reynolds, J.E. \ 2002, Publ. Astron. Soc. Aust., 19, 207

\reference{} Manchester, R.N., Gaensler, B.M., Staveley-Smith, L., Kesteven, M.J., \& Tzioumis, A.K. \ 2005, ApJ, 628, L131

\reference{} Meaburn, J., Bryce, M., \& Holloway, A.J.\ 1995, A\&A, 299, L1

\reference{} McCray, R.\ 2005, in IAU Colloq. 192, Cosmic Explosions, ed. J.M. Marcaide \& K.W. Weiler (Heidelberg: Springer), 77

\reference{} Michael, E., McCray, R., Borkowski, K., Pun, C.S.J., \& Sonneborn, G.\ 1998, ApJ, 492, L143

\reference{} Michael, E., et al. \ 1998, ApJ, 509, L117

\reference{} Michael, E. \ 2000, ApJS, 127, 429.

\reference{} Michael, E., et al. \ 2000, ApJ, 542, L53

\reference{} Michael, E., et al. \ 2002, ApJ, 574, 166

\reference{} Michael, E., et al.\ 2003, ApJ, 593, 809

\reference{} Panagia, N., Gilmozzi, R., Macchetto, F., Adorf, H.-M., \& Kirshner, R.P. \ 1991, ApJ, 380, L23

\reference{} Park, S., Zhekov, S.A., Burrows, D.N., \& McCray, R. \ 2005, ApJ, 634, L73

\reference{} Plait, P.C., Lundqvist, P., Chevalier, R.A., \& Kirshner, R.P. \ 1995, ApJ, 439, 730

\reference{} Pun, C.S.J., et al.\ 2002, ApJ, 572, 906

\reference{} Rousseau, J., Martin. N., Prevot, L., Rebeirot, E., Robin, A., \& Brunet, J.P. \ 1978, A\&AS

\reference{} Scuderi, S., Panagia, N., Gilmozzi, R., Challis, P.M., \& Kirshner, R.P. \ 1996, ApJ, 465, 956

\reference{} Schuster, A. \ 1905, ApJ, 21, 1

\reference{} Shigeyama, T., \& Nomoto, K.\ 1990, ApJ, 360, 242

\reference{} Smith, N., Zhekov, S.A., Heng, K., McCray, R., Morse, J.A., \& Gladders, M. \ 2005, ApJ, 635, L41

\reference{} Sonneborn, G., et al.\ 1998, ApJ, 492, L139

\reference{} Sugerman, B.E.K., Lawrence, S.S., Crotts, A.P.S., Bouchet, P., \& Heathcote, S.R.\ 2002, ApJ, 572, 209

\reference{} Wang, L., \&  Wampler, E.J. \ 1992, A\&A, 262, L9

\reference{} Zhekov, S.A., McCray, R., Borkowski, K.J., Burrows, D.N., \& Park. S. \ 2005, ApJ, 628, L127

\reference{} Zheng, Z., \& Miralda-Escud\'{e}, J. \ 2002, ApJ, 578, 33


\end{references}
\end{document}